\documentclass[aps,pra,twocolumn,superscriptaddress,amsfonts,floatfix,showpacs]{revtex4-1}
\pdfoutput=1
\usepackage{graphicx}
\usepackage{array}
\usepackage{amsmath}
\usepackage{color}

\begin{document}


\title{Analyzing the Toffoli gate in disordered circuit QED}


\author{Jalil Khatibi Moqadam}
\author{Renato Portugal}
\affiliation{Laborat\'orio Nacional de Computa\c{c}\~{a}o Cient\'ifica (LNCC), Rio de Janeiro, Brazil}

\author{Nami Fux Svaiter}
\affiliation{Centro Brasileiro de Pesquisas F\'isicas (CBPF), Rio de Janeiro, Brazil}

\author{Gilberto de Oliveira Corr\^ea}
\affiliation{Laborat\'orio Nacional de Computa\c{c}\~{a}o Cient\'ifica (LNCC), Rio de Janeiro, Brazil}
\date{\today}

\begin{abstract}
We study the effects of imperfections on the fidelity of the Toffoli gate recently realized in a circuit~QED setup using quantum control methods. The noise is introduced in the interqubits interactions. The coupling constants are no longer fixed; instead, they fluctuate around average values obeying some given probability density functions characterizing the dynamical-imperfection case. We also consider the static-imperfection case in which the values of the coupling constants are not exactly known.  We obtain a more robust gate by modifying the quantum optimization problem using a weighted average of the fidelity over an interval of coupling values as the objective functional.
\end{abstract}



\maketitle

\section{Introduction}
Physical implementations of quantum information processing are always subjected to various imperfections that decrease the performance of the process. Dynamical imperfections as a result of system-environment coupling produce decoherence in the system destroying the benefits of using quantum information. Static imperfections on the other hand do not introduce decoherence to the system yet lead to error as well.

A two dimensional lattice of qubits with nearest-neighbor interqubit couplings has already been considered as a standard generic quantum computer model to incorporate imperfections~\cite{quantum_computer_model,quantum_computer_imperfection}. This model shows that, for a system affected by static imperfections, quantum chaos sets in above a critical interqubit strength and annihilates the quantum computer performance~\cite{quantum_computer_imperfection}. Consequently, the entanglement dynamics also exhibits a transition from integrability to quantum chaos~\cite{entanglement_imperfections,multipartite_entanglement_imperfections}. However, the dynamics remains almost unaffected for disorder less than 10\%~\cite{entanglement_dynamics_chains_noise_disorder,connected_network_sc_qubits}. The same model was used to study dynamical imperfections in quantum computers in Ref.~\cite{dynamical_imperfections}, where a characteristic frequency is associated with the noise specifying the rate at which the noise changes. It
was shown that for low frequencies the imperfections can be considered static, and for sufficiently high frequencies the effect of noise completely disappears.

Implementations of quantum computers specifically require high fidelity quantum gates. There are many measures for the robustness of a quantum gate against noise~\cite{robustness_quantum_gates}. In Particular, one- and two-qubit quantum gates that are universal for quantum computing have already been analyzed under the influence of noise~\cite{gate_errors_solid_state,two_qubit_gates_broad_band_noise,noise_filtering_nontrivial_quantum_logic_gates,Modeling_quantum_noise_hardware_quantum_computer}. However, the implementation of multi-qubit gates using standard decompositions in terms of a universal gate set may not be efficient because the implementation time may exceed the decoherence time. It is interesting to analyze efficient ways of implementing multi-qubit gates directly.

The Toffoli gate, a three-qubit gate with a central role in quantum information processing, has been recently implemented with fewer resources and moderate fidelities~\cite{toffoli_gate_trapped_ions,toffoli_gate_superconducting_circuits,three_qubit_qec_sc_circuits}. However, new proposals have been suggested to realize the Toffoli gate with fidelities above 99\%~\cite{toffoli_quantum_control,three_qubit_toffoli_gate_single_step}. Specially in Ref.~\cite{toffoli_quantum_control}, a quantum control scheme has been proposed to realize the Toffoli gate in a circuit quantum electrodynamics (circuit QED) set up. The gate time is about 140 ns that is fast enough because the decoherence time $T_2$ in such systems is 10 to 20 $\mu$s. The usage of quantum control schemes was analyzed in Refs.~\cite{book_dalessandro,quantum_control_unifying_programming_framework,quantum_optimal_control_theory,optimal_control_cqed,local_control_heisenberg_chain,local_control_XXZ} and circuit QED in Refs.~\cite{book_zagoskin,
cqed_architecture,
Coupling_sc_qubits_cavity_bus,cqed_girvin}.

In this paper, we study the effect of imperfections on the Toffoli gate implementation. The Hamiltonian is bilinear with an $XY$-type Heisenberg chain for the system and a Zeeman-like term for the control part. Both parts are subjected to noise. The effect of noise on the control fields has already been considered~\cite{toffoli_quantum_control,local_control_heisenberg_chain,local_control_XXZ} showing that the gate is more robust when the single control pulse duration is reduced. Actually, for a fixed gate time, increasing the noise has less of an effect on the average fidelity of the gate with a higher number of control pulses. However, we investigate the effect of both static and dynamical noise on the interqubit couplings. We also use a control method to find a new set of control fields improving the fidelity affected by noise. The results are based on numerical simulations.

The paper is organized as follows: Sec.~\ref{sec:implementation} is an overview of the quantum control method proposed in Ref.~\cite{toffoli_quantum_control}, which realizes the Toffoli gate in circuit QED. In Sec.~\ref{sec:systems_with_imperfections}, we describe the noise model and analyze the effect of the dynamical and static noise on the
gate fidelity. In Sec.~\ref{sec:improving_robustness}, we obtain a more robust Toffoli gate by modifying the objective functional. Finally, the summary and discussion are presented in Sec.~\ref{sec:summary_discussion}.
\section{\label{sec:implementation} Implementation of the Toffoli Gate}
Transmon qubits can be coupled together within circuit QED through a transmission line resonator. They can be isolated from electromagnetic environment and controlled through resonant microwave drives~\cite{transmon}. Quantum control techniques can then be applied in this setup to implement various quantum information processing tasks~\cite{book_dalessandro}.

The interaction Hamiltonian of an array of three transmon qubits coupled to a superconducting transmission-line resonator can be effectively described by an $XY$-type (flip-flop) Hamiltonian~\cite{toffoli_quantum_control,map_JC_XY}
\begin{equation}
 \label{eq:XY-type}
 H_{0}=\sum_{i<j}J_{ij}\left({\sigma}_{ix}{\sigma}_{jx}+{\sigma}_{iy}{\sigma}_{jy}\right),\;\;\;\;\;i,j=1,2,3
\end{equation}
where $J_{ij}$ are coupling constants and ${\sigma}_{ix}$ and ${\sigma}_{iy}$ are Pauli $X$ and $Y$ matrices, respectively. This system is manipulated by a Zeeman-like Hamiltonian
\begin{equation}
 \label{eq:zeeman}
 H_{c}(t)=\sum_{i=1}^{3}[{u^{(i)}_{x}(t)} {\sigma_{ix}}+{u^{(i)}_{y}(t)}{\sigma_{iy}}],
\end{equation}
with control fields $u^{(i)}_{x}(t)$ and $u^{(i)}_{y}(t)$ affecting the qubits in $x$ and $y$ directions. Therefore, the system dynamics is governed by the total Hamiltonian
\begin{equation}
 H(t)=H_0+H_c(t).
 \label{eq:total_hamiltonian}
\end{equation}

These control fields can be implemented through wave generators. However, to keep the transmon qubits well-defined two-level systems, the fields cannot be arbitrarily large. The norm
\begin{equation}
 u_{\max}=\max_{i,t}\sqrt{ [ u^{(i)}_{x}(t)]^2+[u^{(i)}_{y}(t)]^2}
\end{equation}
is then restricted to be smaller than some threshold value.

The controllability of the system can be verified by considering the Lie algebra generated by $span_{j=1,2,3}\{-iH_0,-i \sigma_{jx},-i \sigma_{jy}\}$ which is actually the Lie algebra $su(8)$. This algebra provides the operator controllability of the system~\cite{book_dalessandro}.

Finding the control fields that implement the Toffoli gate is a numerical optimization problem. A given gate time $t_g$ is divided into $N_t$ (even number) intervals of the same time length $T$. Then, the set of fields is supposed to be constant in each interval and is acting on the related qubits alternating between the $x$ and $y$ directions.

In the first interval $0\leq t\leq T$, three $x$-control pulses with constant amplitudes $u_{x,1}^{(i)}$ are applied to all three qubits. The dynamics is then governed by the Hamiltonian $$H_{x,1}=H_0+\sum_{i=1}^{3}u_{x,1}^{(i)}\sigma_{ix}.$$ In the second interval $T \leq t \leq 2T$, the $y$-control pulses with amplitudes $u_{y,1}^{(i)}$ are applied leading to the Hamiltonian $$H_{y,1}=H_0+\sum_{i=1}^{3}u_{y,1}^{(i)}\sigma_{iy}.$$ This process is repeated $N_t/2$ times to complete all $N_t$ intervals. These time-independent Hamiltonians for each interval lead to the time evolution operators $U_{x,n}=\exp(-iH_{x,n}T)$ and $U_{y,n}=\exp(-iH_{y,n}T)$ respectively in corresponding intervals. The product $U_{y,n}U_{x,n}$, for $n=1,\dots,N_t/2$ in the reverse order, is the time evolution operator $U(t=t_g)$.

The values of the $3N_t$ control fields are obtained through maximizing the fidelity
\begin{equation}
\label{eq:fidelity}
F=\frac{1}{8}\left|\;\mathrm{Tr}\left[U^\dagger\left(t_g,N_t,\mathbf{u},\{J_{ij}\}\right)U_{\mathrm{Toff}}\right]\;\right|,
\end{equation}
where $\mathbf{u}$ is the concatenation of all control pulses and $U_{\mathrm{Toff}}$ is the Toffoli gate, given by
\begin{equation}
	 U_{\mathrm{Toff}}=\begin{bmatrix}
    1\ 0\ 0\ 0\ 0\ 0\ 0\ 0 \\
    0\ 1\ 0\ 0\ 0\ 0\ 0\ 0 \\
    0\ 0\ 1\ 0\ 0\ 0\ 0\ 0 \\
    0\ 0\ 0\ 1\ 0\ 0\ 0\ 0 \\
    0\ 0\ 0\ 0\ 1\ 0\ 0\ 0 \\
    0\ 0\ 0\ 0\ 0\ 1\ 0\ 0\\
    0\ 0\ 0\ 0\ 0\ 0\ 0\ 1 \\
    0\ 0\ 0\ 0\ 0\ 0\ 1\ 0 \\
  \end{bmatrix}
\end{equation}
in the computational basis. The local optimized control fields are obtained through numerical maximization of the fidelity with an initial guess for the fields. The minimum gate time for a given fidelity can also be obtained by starting with a sufficiently large $t_g$ and then gradually decreasing its value.

The piecewise-constant control pulses obtained here can be filtered through a low-pass filter such that they can be generated by an actual wave generator. Using a product formula approach (see Ref.~\cite{local_control_heisenberg_chain}), it is possible to obtain the fidelities for the filtered control fields.

Assuming $u_{\max}<130$ MHz, $J=30$ MHz and \linebreak$J_{12}=J_{23}=6J_{13}=J$, the Toffoli gate can be realized with a fidelity larger than 99\% in $4.18J^{-1}=140$ ns with $N_t=20$ control intervals. In the following sections we focus on this specific realization.
\section{\label{sec:systems_with_imperfections} The Toffoli Gate in Systems with Imperfections}
In this section we investigate how the gate fidelity is affected by the noise in the system. To do so the system Hamiltonian Eq.~(\ref{eq:XY-type}) is subjected to noise while the control Hamiltonian Eq.~(\ref{eq:zeeman}) is maintained perfect (without noise). The scenario is to apply a set of control fields optimized for the perfect system to an imperfect system. In other words, we analyze the sensitivity of the control fields to the noise in the main system.
\subsection{\label{sec:noise_model}Noise Model}
We are especially interested in the case where the interactions between the qubits are noisy. Each $J_{ij}$ in Eq.~(\ref{eq:XY-type}) is considered to be independently coupled to a stochastic variable described by
\begin{equation}
 J_{ij}(t)=\bar{J}_{ij} \left( 1+\epsilon_{ij}(t) \right),
 \label{eq:time_dependent_coupling}
\end{equation}
where $\bar{J}_{ij}$ are average coupling strengths and $\epsilon_{ij}(t)$ are independent Gaussian random variables. This noise model was used in many physical contexts and can be seen as an effective behavior when we average out the effect of the environment over the main system~\cite{spin_glass}.

We suppose that all $\epsilon_{ij}(t)$ have the same mean and variance and change their values in time simultaneously with a fixed frequency $f_c$. Then, they remain fixed during the time interval $\tau_c=1/f_c$:
\begin{equation}
 \epsilon(t) = \epsilon^{(k)},\;\;\;\;\;(k-1)\tau_c \leq t < k \tau_c,
 \label{eq:sinc2_noise}
\end{equation}
where we have omitted the subindices because all components have the same behavior. They are independent and identically distributed (iid) Gaussian random variables in different time intervals with expectation $\mathrm{E} [ \epsilon^{(k)} ]=0$ and variance $\mathrm{E} [ (\epsilon^{(k)})^2 ] = \sigma^2 < \infty$, where $k$ runs over integer values.

The autocorrelation function for $\epsilon(t)$ is given by
$$\langle \epsilon(t) \epsilon(t+\tau) \rangle = \left(1 - \frac{\left| \tau \right| }{ \tau_c}\right)\sigma^2$$
for $\tau$ in $[-\tau_c,\tau_c]$ and zero otherwise. The power spectral density (Fourier transform of the autocorrelation function), which displays the essence of the noise, is
\begin{equation}\label{eq:sinc}
    S(\omega) = \frac{\sigma^2\tau_c }{\sqrt{2\pi}}\,\mathrm{sinc}^2\left(\frac{\omega \tau_c}{2}\right).
\end{equation}

In Sec.~\ref{sec:dynamic} we describe how we simulate the effect of noise on the fidelity for values of $\tau_c$ such that $\tau_c \leq t_g$. For $\tau_c > t_g$ we are only interested in the limit $\tau_c\rightarrow \infty$. In this case the power spectral density function approaches to the delta function on zero, which corresponds to a fixed noise in the system. This means that the value of the random variable remains fixed for ever. Physically, this situation is associated with inaccuracies in the system parameters. This situation is analyzed in Sec.~\ref{sec:static}.
\subsection{\label{sec:dynamic}Dynamical Imperfections}
We start by obtaining a set of optimal control fields implementing the Toffoli gate in the perfect system. The fields can be found by maximizing the fidelity given by Eq.~(\ref{eq:fidelity}) using the parameter values given at the end of Sec.~\ref{sec:implementation}. To be close enough to the global optimal solution, we carry the maximization process over 200 random initial guesses and then select the set of fields with the largest fidelity.

In our noisy system, the couplings $J_{ij}$ (see Eq.~(\ref{eq:XY-type})) remain no longer fixed and evolve according to Eq.~(\ref{eq:time_dependent_coupling}). There are three different coupling constants and, in principle, the noise affects each of them independently. We are imposing that $\bar{J}_{12}=\bar{J}_{23}=6\bar{J}_{13}=\bar{J}$ and we are assuming that the random variables $\epsilon_{ij}(t)$ associated with those couplings have the same mean, variance and characteristic frequency. Under those assumptions, we will show later on in this section that introducing the noise only in $J$ is essentially equivalent to introducing the noise in the three coupling constants. This explains why we can drop the subindices of $\epsilon_{ij}(t)$ and denote the random variable simply by $\epsilon(t)$.

Let $\epsilon(t)$ change its value with a fixed frequency $f_c=1/\tau_c$. We generate $t_g f_c$ random numbers independently according to a Gaussian distribution with a zero mean and standard deviation $\sigma$. Accordingly, we have a realization of $\epsilon(t)$ and therefore $J(t)$ for the whole interval $[0,t_g]$. It is now possible to calculate the time evolution operator using the total Hamiltonian $H_0(J(t))+H_c(t)$ with those values of $J(t)$ realized above. The corresponding fidelity for that realization can then be calculated using Eq.~(\ref{eq:fidelity}).

The next step is to calculate the average fidelity, which is obtained by repeating the above process with the same $f_c$ and $\sigma$ for a large number of realizations of $\epsilon(t)$ within the corresponding time interval and summing over all fidelities and dividing by the total number of realizations. Finally, using this method systematically, we obtain the fidelity as a function of $\sigma$ with a fixed $f_c$.

\begin{figure}
\includegraphics[trim = 17mm 68mm 20mm 65mm, clip=true, width=9cm]{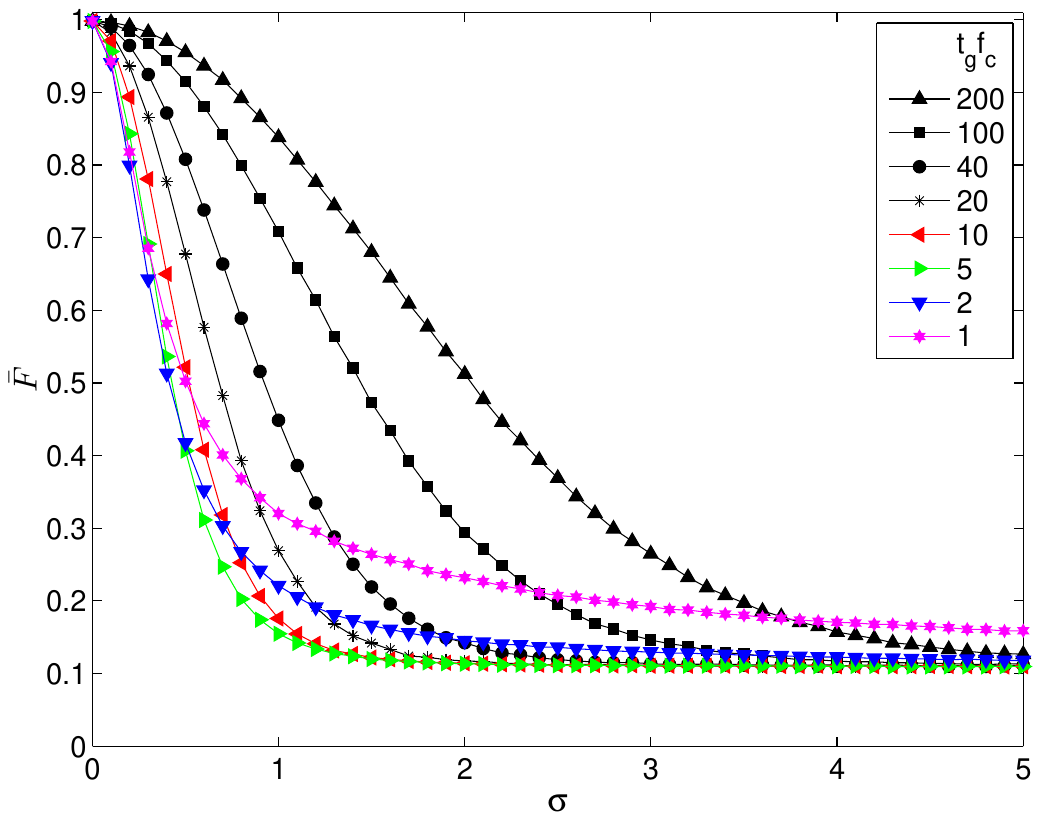}
\caption{\label{fig:dynamicfidversigma}(Color online) The average fidelity versus standard deviation $\sigma$ for $t_g f_c = 1,2,5,10,20,40,100,200$. The set of control fields used in each case has been obtained by maximizing Eq.~(\ref{eq:fidelity}), which leads to a fidelity about 99.83\% with respect to the perfect system.}
\end{figure}

Figure~\ref{fig:dynamicfidversigma} shows the average fidelity as a function of $\sigma$ for $t_g f_c = 200,100,40,20,10,5,2,1$ from top to bottom, respectively. The number of realizations per each fixed $\sigma$ for the first four items is 10,000 and for the last four items is 100,000. The average fidelity decreases for all frequencies when the standard deviation $\sigma$ increases, confirming our intuition about the effect of the decoherence over the coherent evolution. Notice that the average fidelity quickly drops to small values specially when $t_gf_c=5$.

For a given $\sigma$, the fidelity is less affected when the noise characteristic frequency is high ($t_gf_c\gg1$). In other words, the high-frequency noise generates less decoherence. This observation in our simulations can be justified first of all in terms of our noise model specially by looking at the power spectral density given by Eq.~(\ref{eq:sinc}). As the noise frequency increases $(\tau_c \rightarrow 0)$, the power spectral density function approaches zero for all $\omega$. This means that the noise disappears and the fidelity remains unaffected.

However, an alternative way to analyze this result is using the theorem given in Ref.~\cite{dynamical_imperfections}. The time evolution of the system over each control time $T=\tau_cN$ is given by
\begin{equation}
 U_N(T)=\prod_{k=1}^{N}\exp[-i(\bar{H}+\epsilon^{(k)} \bar{H}_0)\tau_c],
 \label{eq:evolution_noisy_control_time}
\end{equation}
where $\bar{H}_0$ and $\bar{H}$ are obtained from $H_0$ and $H$, Eq.~(\ref{eq:XY-type}) and  Eq.~(\ref{eq:total_hamiltonian}) respectively, after replacing $ \{ J_{ij} \}$  by $\{ \bar{J}_{ij} \}$. Expanding the product, we can write the result up to the first power of $T$ as 
$$1 - i\bar{H}T -  \frac{i\bar{H}_0T}{N} \sum_{k=1}^{N} \epsilon^{(k)}.$$ 
But according to the weak law of large numbers
$$\lim_{N \rightarrow \infty}  \left( \left|\frac{1}{N}\sum_{k=1}^{N}\epsilon^{(k)} - E[\epsilon^{(k)}]\right| > \varepsilon \right) = 0, $$
and thus the first power of $\bar{H}_0T$ as well as all the higher powers approach zero in probability, hence
\begin{equation}\label{eq:limUN}
 \lim_{N \rightarrow \infty} \left( \parallel U_N(T) - \exp(-i\bar{H}T) \parallel > \varepsilon \right) = 0.
\end{equation}
Therefore, the effect of noise disappears if the characteristic frequency $\tau_c^{-1}$ is sufficiently high. 

The validity of Eq.~(\ref{eq:limUN}) can also be verified by decreasing the number of realizations, that is, performing an average with smaller samples. In this case, the resulting plots will no longer be smooth. Actually, the points in each plot are scattered around the average curve. However, the smoothness of the plots corresponding to high frequencies is less affected. In another words, the plots for higher frequencies are less sensitive to the number of realizations.

However, using the central limit theorem,
$\eta=\lim_{N \rightarrow \infty} \sum_{k=1}^{N}\epsilon^{(k)}/\sqrt{N}$ goes to a Gaussian distribution with the same mean and variance of $\epsilon^{(k)}$. Therefore, for finite but large $N$~\cite{dynamical_imperfections}
\begin{equation}
  U_N(T) \sim \exp(-i\bar{H}T) \exp(-i \eta \bar{H}_0 T / \sqrt{N}),
  \label{eq:central_limit}
\end{equation}
which shows how the error enters in each control step $T$ of the implementation and the errors will eventually reduce the gate fidelity.
This result is not valid for intermediate values of $N$, since the terms corresponding to the nonzero commutator brackets of $\bar{H}_0$ and $\bar{H}$ become also important.

Fig.~\ref{fig:dynamicfidversigma} also shows that for a fixed $\sigma$ the average fidelity decreases when we decrease the noise frequency from $t_g f_c = 200$ until $t_g f_c = 5$ and the fidelity has the opposite behavior from $t_g f_c =5 $ until $t_g f_c =1$. This behavior can be explained from the effect of the control pulses on an imperfect system. When the control-pulse frequency is sufficiently larger than the noise frequency its effect helps to protect the system against decoherence, similar to dynamical decoupling~\cite{dynamical_decoupling_viola,dynamical_decoupling_lidar}. When we increase the noise frequency, the control-field effect becomes less relevant, and for $t_g f_c > 5$ it plays almost no role at all.

Finally, Fig.~\ref{fig:dynamicfidversigma} shows that the curves apparently converge to the same average fidelity for sufficiently large $\sigma$. The limiting average fidelity is close to ${1}/{d}=0.125$, which corresponds to the one obtained from the full randomizing map $\varepsilon(\rho)={\mathbb{1}}/{d}$, where $\rho$ is an arbitrary (pure) state and $d$ is the Hilbert space dimension. However, this saturation value (0.125) can be obtained if the randomness in the gate parameters uniformly generates all elements in $SU(8)$, and if we end up in a Lie subgroup of $SU(8)$, the saturation value can be different (see Ref.~\cite{local_control_heisenberg_chain}). What we observe after performing simulations with large values of $\sigma$ is that the saturation values are not exactly the same, but slightly different depending on the noise frequency. Therefore, fixing the control Hamiltonian and just changing the coupling $J$ does not seem to generate the whole $SU(8)$ space.

Now, we come back to justifying why introducing one source of noise to $J$ is sufficient to obtain essentially the same results of having three independent sources of noise affecting three couplings $\{J_{1 2},J_{1 3},J_{2 3}\}$. In the latter case, the exponent in Eq.~(\ref{eq:evolution_noisy_control_time}) is replaced by
\begin{equation*}
 -i( \bar{H} + \boldsymbol{\epsilon}^{(k)} \cdot \mathbf{\bar{H}}_0)\tau_c,
\end{equation*}
where $\boldsymbol{\epsilon}^{(k)} = (\epsilon_{12}^{(k)},\epsilon_{13}^{(k)},\epsilon_{23}^{(k)})$, $\mathbf{\bar{H}}_0 = (H_{12},H_{13},H_{23})$, and $$H_{ij} = \bar{J}_{ij} \left({\sigma}_{ix}{\sigma}_{jx}+{\sigma}_{iy}{\sigma}_{jy}\right),\,i<j.$$ By using $\boldsymbol{\eta}=\lim_{N \rightarrow \infty} \sum_{k=1}^{N}\boldsymbol{\epsilon}^{(k)}/\sqrt{N}$ instead of $\eta$, a similar reasoning to the one used for one $J$ can be applied here to reach the same results as before for high noise frequencies. Moreover, we have repeated the simulations with three independent sources of noise and found essentially the same results for low noise frequencies as well. Actually, increasing the number of noise sources just leads to faster decay of average fidelity for all frequencies. Simulations with six different couplings (different couplings in the $X$ and $Y$ directions) and with six independent sources of noise confirm the latter statement. In the new simulations, there is no universal saturation value.

\subsection{\label{sec:static}Static Imperfections}
As we discussed at the end of Sec.~\ref{sec:noise_model}, the case $\tau_c \rightarrow \infty$ corresponds to a fixed noise in the system, that is, the noise does not change in time. Such a static noise can be associated with, for example, inaccessibility of measuring exactly the system parameters such as the coupling constants. Static noises do not lead to any decoherence, but they introduce error into the gate implementation. It is known that entanglement dynamics in similar systems remains almost unchanged under the influence of uniform static noise when it is smaller than 10\% (see Refs.~\cite{entanglement_dynamics_chains_noise_disorder,connected_network_sc_qubits}).

Analyzing the error for the static-imperfections case is mathematically equivalent to the case of dynamical noise with $\tau_c^{-1}=t_g^{-1}$. Again, we adjust the control Hamiltonian for the perfect system and use it for a system with static imperfections. The coupling is given by Eq.~(\ref{eq:time_dependent_coupling}) with a fixed random variable $\epsilon(t)=\epsilon$ for the whole evolution. Considering a uniform distribution for $\epsilon$, we generate a large sample of couplings with a given half-width $\delta$ and then calculate the corresponding fidelity using Eq.~(\ref{eq:fidelity}). After obtaining the sample average and repeating the procedure for many values of $\delta$, we find how the fidelity changes as a function of the half-width. Moreover, we have decreased the noise level interval to be more focused on the region with unaffected entanglement.

\begin{figure}
\includegraphics[trim = 15mm 68mm 20mm 65mm, clip=true, width=9cm]{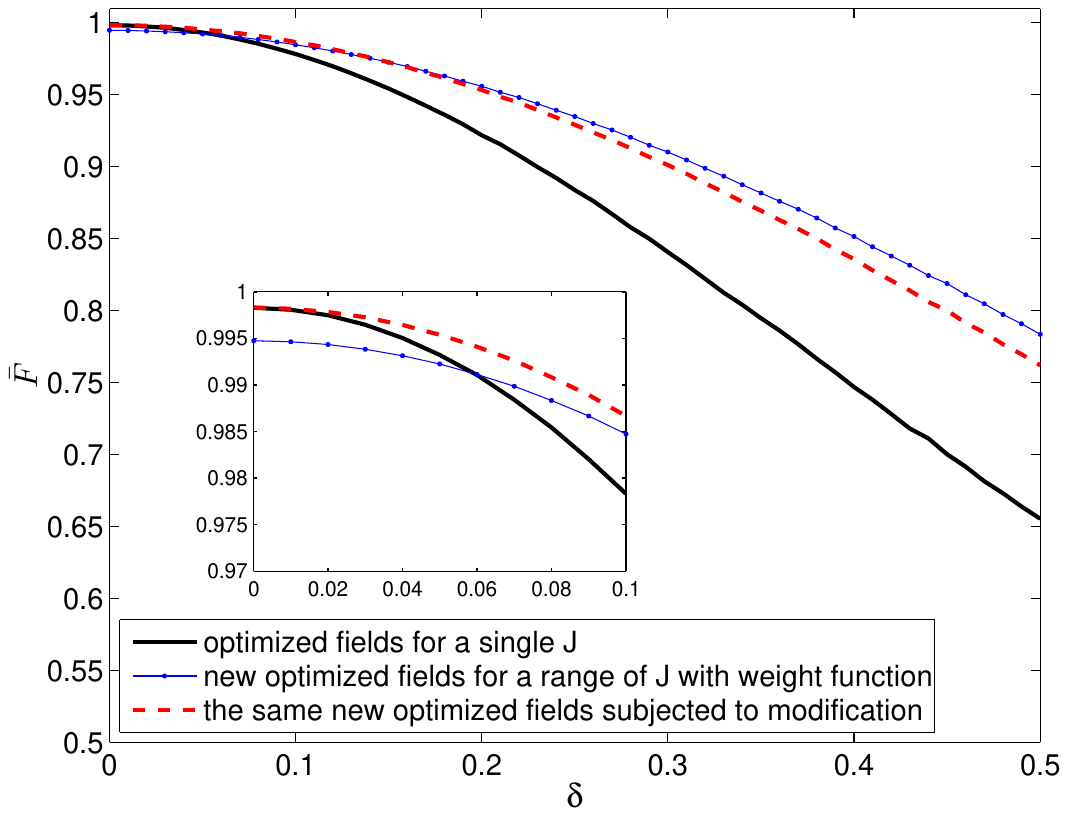}
\caption{\label{fig:staticfidversigma}(Color online) The average fidelity versus half-width $\delta$ (static imperfections) for three different sets of control fields: the solid line corresponds to the fields obtained by maximizing Eq.~(\ref{eq:fidelity}), the dotted line corresponds to the fields obtained by solving the new optimization problem~(\ref{eq:newobjective}) with the weight function~(\ref{eq:weight_function}), and the dashed line shows the result after solving the same problem but with the weight function~(\ref{eq:inverse_weight_function}).  The corresponding fidelities are about 99.83\%, 99.47\% and 99.83\% for the perfect system, respectively. Inset: a zoom on the region $0 \leq\delta \leq 0.1$.}
\end{figure}

The solid curve in Fig.~\ref{fig:staticfidversigma} depicts the average fidelity as a function of the half-width $\delta$. The number of realizations per each fixed $\delta$ is 100,000. The diagram shows that the average fidelity decreases when the half-width $\delta$ increases. Here, we have focused on half-widths less than 0.5 because the experimental upper bound for such a superconducting qubit chain is quite a bit below that~\cite{entanglement_dynamics_chains_noise_disorder}. Actually, for disorders such as $\delta \leq 10\%$, the average fidelity remains above 97.83\%. Other curves in Fig.~\ref{fig:staticfidversigma} will be discussed in the next section.

\section{\label{sec:improving_robustness}Improving the Robustness}
In Sec.~\ref{sec:systems_with_imperfections}, we have considered the set of control fields optimized for a fixed value of coupling and then applied that set to a system with imperfections. Actually, we have investigated there the performance of that specific set of fields in the presence of noise. However, it is possible to find another set of control fields that are more robust to noise. In this section, we use the optimal control techniques to find a different set of control fields which yields improved performance in systems with imperfections.

The set of control fields which is used in Sec.~\ref{sec:systems_with_imperfections} is the solution $\mathbf{u}$ of the optimization problem
\begin{equation}
  \max_{ \mathbf{u} }{ F\left( \mathbf{u} , \bar{J} \right) },
 \label{eq:old_optimization}
\end{equation}
where $F$ is given by Eq.~(\ref{eq:fidelity}). But, the objective functional $F \left( \mathbf{u} , \bar{J} \right)$ uses a single coupling value $J=\bar{J}$, and therefore the resulting optimal solution has the best fidelity at that specific coupling irrespective of the fidelity at points for which $J\neq\bar{J}$. However, with noisy couplings a more robust set of control fields should have high fidelities also in the couplings that deviate from $J=\bar{J}$. Toward that end, we modify the objective functional such that it includes a range of coupling values $[\bar{J}-\delta J,\bar{J}+\delta J]$. We solve the new optimization problem
\begin{equation}
  \max_{ \mathbf{u} }{ \int_{\bar{J}-\delta J}^{\bar{J}+\delta J}F(\mathbf{u},J)w(J)\,dJ},
 \label{eq:newobjective}
\end{equation}
where $w(J)$ is an appropriate weight function enabling us to put different stress on the points in the interval brought into the process of optimization. Actually, the objective functional in problem~(\ref{eq:newobjective}) is a weighted average of the fidelity over an interval of coupling values.

\begin{figure}
\includegraphics[trim = 15mm 68mm 20mm 65mm, clip=true, width=9cm]{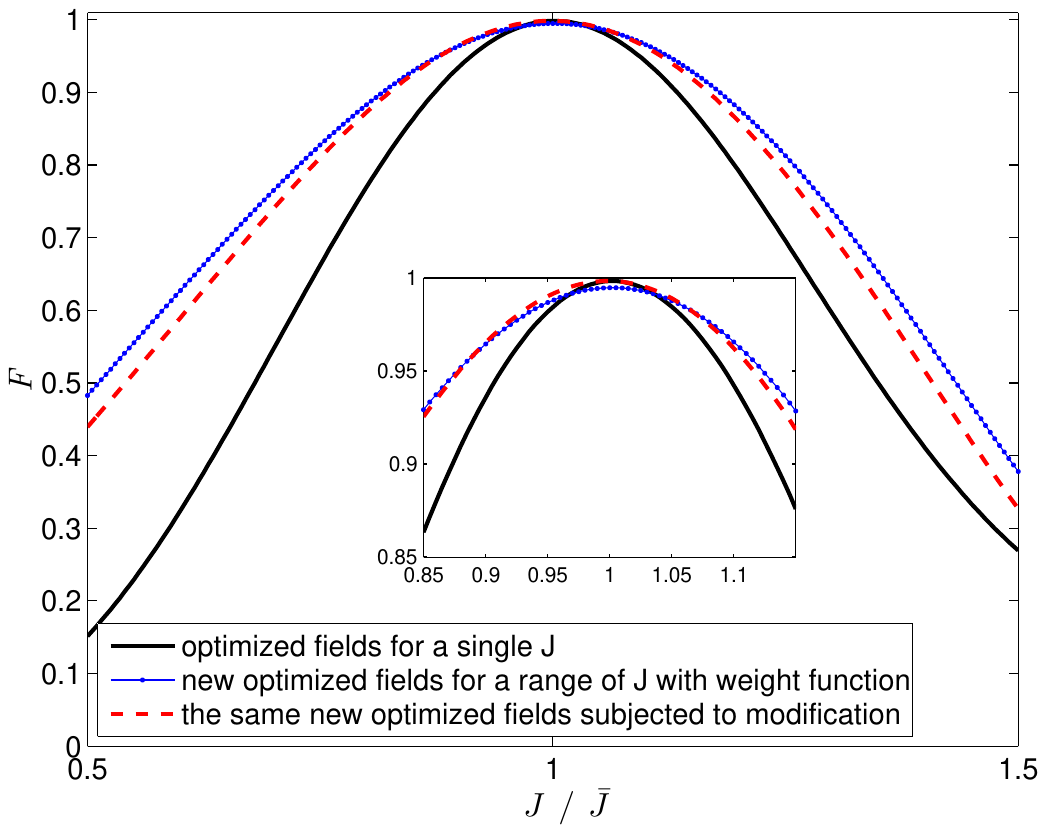}
\caption{\label{fig:robustfidverj418}(Color online) Fidelity versus $J/\bar{J}$ for three different sets of control fields: the solid line corresponds to the fields obtained by maximizing Eq.~(\ref{eq:fidelity}), the dotted line corresponds to the fields obtained by solving the new optimization problem~(\ref{eq:newobjective}) with the weight function~(\ref{eq:weight_function}), and the dashed line shows the result by solving the same problem but with the weight function~(\ref{eq:inverse_weight_function}). Inset: a zoom on the peak.}
\end{figure}

The solid line in Fig.~\ref{fig:robustfidverj418} depicts the fidelity in terms of $J/\bar{J}$ for the fields optimized for a single value of coupling $J=\bar{J}$ obtained from solving problem~(\ref{eq:old_optimization}). Inspired by this plot, we set the weight function such that it has small values in the vicinity of $J/\bar{J}=1$ and large values in the outermost points of the interval. In fact, by simply letting
\begin{equation}
 w(J)= \left\{ \begin{array}{ll}
              0, \;\;\;\;\; & \left| \frac{J}{\bar{J}}-1 \right| \leq \delta_1\\
              \;\\
              1, \;\;\;\;\; \delta_1  < & \left| \frac{J}{\bar{J}}-1 \right| \leq \delta_2,
              \end{array} \right.
              \label{eq:weight_function}
\end{equation}
we can find a new set of control fields leading to higher fidelities in points different from $J/\bar{J}$. Solving the new optimization problem with 200 random initial guesses and choosing the fields with highest fidelity, we can approach the global solution in this case. The dotted line in Fig.~\ref{fig:robustfidverj418} depicts the corresponding fidelity in terms of $J/\bar{J}$ for such a set of fields with $\delta_1=0.05$, $\delta_2 = 0.15$, and $\delta J = 0.15 \bar{J}$. It can be seen that the new optimal fields, compared with the previous ones, have smaller fidelities in the vicinity of $J/\bar{J}=1$ but larger in other points.

However, we can improve the fidelities in the vicinity of $J/\bar{J}=1$ by setting
\begin{equation}
w(J)=F(\mathbf{u}_{0},J)^{-1},
\label{eq:inverse_weight_function}
\end{equation}
where $\mathbf{u}_0$ is the global solution of the problem~(\ref{eq:newobjective}) with the weight function~(\ref{eq:weight_function}). The dashed line in Fig.~\ref{fig:robustfidverj418} shows the corresponding fidelity versus $J/\bar{J}$ for such a set of fields with $\delta J = 0.1 \bar{J}$ and the initial guess equal to $\mathbf{u}_0$.

Considering Fig.~\ref{fig:robustfidverj418}, we see that the dotted line has the best worst-case performance
having the largest values at the boundaries $J/\bar{J}=0.5,1.5$. However, in the
narrower interval $[0.9,1.1]\bar{J}$ the dotted line and the dashed line have almost the same
worst-case performance but the dashed line is above the dotted line in a larger part of the interval. As
described below, among the three sets of control fields discussed here, the dashed line corresponds to the fields that have the best performance in systems with static disorder less than~10\%.

The performance of our new sets of control fields in the presence of noise can be analyzed in the same way as in Sec.~\ref{sec:systems_with_imperfections}. In the case of static imperfections, the result for the fields optimized according to problem~(\ref{eq:newobjective}) with the weight function~(\ref{eq:weight_function}) is shown by a dotted line in Fig.~\ref{fig:staticfidversigma}. The dashed line in the same figure shows the situation for the fields optimized with the weight function~(\ref{eq:inverse_weight_function}). In particular, for 10\% disorder ($\delta=0.1$) the corresponding fidelities are 98.47\% and 98.67\%, respectively, showing 0.64\% improvement in the first case and 0.84\% improvement in the second case, compared to 97.83\% (original problem~(\ref{eq:old_optimization})).

For the dynamical-imperfections case, when the noise frequency is high, the resulting plots for the three sets of control fields remain almost the same because the effect of control pulses plays no role in high noise frequencies. However, for low noise frequencies $t_g f_c = 5,2,1$ (see Sec.~\ref{sec:dynamic}) the new sets of fields have better performances.

\section{\label{sec:summary_discussion}Summary and Discussion}
In this paper we considered the effect of imperfections on a recently established Toffoli gate realized in circuit QED with quantum control methods. The total Hamiltonian is bilinear and we studied the effect of imperfections on the system Hamiltonian by introducing noise in the interqubit couplings. We showed that in the case of dynamical imperfections the average fidelity is less sensitive to noise for high characteristic frequencies. Actually the effect of noise completely disappears when the noise frequency is sufficiently high. 
For static noise we showed that the fidelity decreases by 2\% when the system is affected by a uniform noise with half-width less than 10\%.

We also obtained two new sets of control fields by modifying the objective functional in the original optimization problem considered in Ref.~\cite{toffoli_quantum_control}. We showed that these new sets of control fields are more robust when affected by static noise.

It may be stressed that the results in Ref.~\cite{toffoli_quantum_control}
as well as those in the present paper
are valid under the two-level approximation, i.e., in the absence of a
significant leakage from the two-state computational subspace of the
transmon qubits. However, the parameter imperfections studied
here can likely be incorporated in an analogous fashion within a more
complete multi-level analysis.

The 2\% reduction of the average fidelity is expected to be also
valid for other three-qubit gates such as the Fredkin gate,
when the system is affected by a uniform noise with half-width less than 10\%.

We considered independent distributions in different time intervals for random variables in the noise model Eq.~(\ref{eq:sinc2_noise}). It is possible to extend this method simulating other types of noise by considering various dependencies between the random variables in different times. Consequently, the average fidelity may be improved in a more effective way having those sorts of dependencies. The techniques used in this work may be applied to other multi-qubit gates as well.

\begin{acknowledgments}
We would like to thank Stefan Boettcher and Hamed Saberi for their stimulating discussions and Ali Rezakhani for the useful comments and directions.
\end{acknowledgments}

\end{document}